\documentclass[twocolumn,showpacs,pre,floatfix]{revtex4}
\usepackage[pdftex]{graphicx}
\usepackage{epstopdf}
\usepackage{epsfig}
\usepackage{color}

\usepackage{amssymb,amsmath,stmaryrd,tabularx}
\SetSymbolFont{stmry}{bold}{U}{stmry}{m}{n}
\usepackage{wrapfig}
\usepackage{bm}

\newcommand{\be}{\begin{equation}}
\newcommand{\ee}{\end{equation}}
\newcommand{\bea}{\begin{eqnarray}}
\newcommand{\eea}{\end{eqnarray}}

\newcommand{\ud}{\mathrm{d}}

\newcommand{\ddt}[2]{\frac{\mathrm{d^2}#1}{\mathrm{d}#2^2}}
\newcommand{\ddx}[2]{\ddt{}x}
\newcommand{\ddy}[2]{\ddt{}y}
\newcommand{\ddz}[2]{\ddt{}z}

\begin{document}
\title{Velocity statistics for non-uniform configurations of point vortices}
\author{Audun Skaugen and Luiza Angheluta}
\affiliation{
Department of Physics, \\ University of Oslo, P.O. 1048 Blindern, 0316 Oslo, Norway}


\begin{abstract}
Within the point vortex model, we compute the probability distribution function of the velocity fluctuations induced by same-signed vortices scattered within a disk according to a fractal distribution of distances to origin $\sim r^{-\alpha}$. We show that the different random configurations of vortices induce velocity fluctuations that are broadly distributed, and follow a power-law tail distribution, $P(V)\sim V^{\alpha-2}$ with a scaling exponent determined by the $\alpha$ exponent of the spatial distribution. We also show that the range of the power-law scaling regime in the velocity distribution is set by the mean density of vortices and the exponent $\alpha$ of the vortex density distribution. 
\end{abstract}
\maketitle
\date{}

\section{Introduction} 
Two-dimensional (2D) turbulent flows are known to exhibit an \emph{inverse energy cascade},  where the kinetic energy is transferred from smaller to larger scales, that can lead to a~\lq condensation\rq~of energy at low-wavenumbers~\citep{kraichnan1967inertial}. This results in the formation of transient, large-scale rotating structures by an aggregation process or clustering of same-sign vortices by analogy to the Richarsdon's cascade of energy by breakdown of vortices in three dimensional turbulence~\citep{tabeling2002two}. For this reason, the vorticity dynamics is crucial in 2D turbulence. Onsager's point vortex model~\citep{Onsager_1949} provides an approximate statistical description of turbulence, where vorticity is represented as a set of localized point vortices described by a Hamiltonian dynamics that generates dynamical regimes of clustering of same-signed vortices. Since the experimental realization of quantum turbulence in 2D Bose-Einstein condensates (BECs)~\citep{neely2013characteristics}, the point-vortex model has become a particularly way powerful to study both the statistical properties of interacting quantized vortices, as well as the analogy between the classical and quantum turbulent cascades~\citep{Bradley_2012,Reeves_2013,Reeves_2014,Skaugen15}. 

Several studies on the clustering regime of same-signed vortices show that an inverse energy cascade develops for a self-similar spatial distribution of clustered vortices where the distance between vortices inside a cluster follows a power-law distribution $\sim r^{-\alpha}$, where $\alpha=1/3$ corresponds to the Kolmogorov scaling of the incompressible energy spectrum in the wavenumber space $E(k)\sim k^{-5/3}$ ~\citep{Novikov_1975,Bradley_2012}.

In a recent numerical study of 2D quantum turbulence~\citep{Skaugen15}, we 
investigated the relationship statistical properties of vortices and the inverse energy cascade, using the damped Gross-Pitaevski equation. In particular, we showed that the vortex clustering regime, contributing to the formation of $k^{-5/3}$ scaling in the incompressible energy spectrum at lenghscales above the mean vortex distance, is also signaled by a power-law tail in the probability distribution of vortex velocity fluctuations $P(V)\sim V^{-5/3}$. However, in these kind of numerical studies, the scaling range both in the energy spectrum and velocity probability distribution is limited by finite-size effects and mean vortex density, and it is numerically challenging to reduce these effects. Therefore, the aim of this paper is to provide an analytical calculation of the $P(V)$ for clustered vortices where we can vary the mean-vortex distance and the system size. This way, we are able to show that the range of the power-law tail increases with decreasing the mean density of clustered vortices, while it is relative robust to finite-size effects related to low number of cluster vortices within a small radius as long as the mean density is kept fixed. Moreover, power-law  exponent for $P(V)$ is directly related to the fractal distribution of clustered vortices as $P(V)\sim V^{\alpha-2}$.

The statistical distribution of velocity fluctuations induced by uniformly distributed random vortex configurations is known to follow $P(V)\sim V^{-3}$ that can be predicted from the point-vortex model~\citep{Chavanis_2000}. Using similar analytical technique~\cite{chandrasekhar1943stochastic}, we calculate the general velocity distribution induced by non-uniform random vortex configurations that have a fractal distribution $r^{-\alpha}$.

\section{Point vortex model for a cluster}
We consider an ensemble of $N$ identical point vortices distributed within a disk 
of radius $R$, such that the probability of having a vortex at position $\vec r$ from the disk's origin follows a power-law with the distance, $\tau(\vec r)\ud \vec r \propto |\vec r|^{-\alpha-1}\ud \vec r$~\citep{Novikov_1975}. Hence, the probability distribution of the \emph{distance from the origin} $r = |\vec r|$ will
then pick up a factor $2\pi r$ from the 2D measure, so that 
\begin{equation}
T(r) \propto 2\pi r r^{-\alpha-1} \propto r^{-\alpha},
\end{equation}
with $\alpha = 1/3$ corresponding to the distribution of clustered vortices that contribute to the inverse energy cascade in two-dimensional quantum turbulence~\citep{Novikov_1975, Bradley_2012}.

Because vortices of opposite sign annihilate when they get within a distance $a$ related to the coherence length $\xi$, we modify the distribution $\tau(\vec r)$ by introducing a lower cutoff $a \sim \xi$ coming from finite vortex cores. Using the normalization condition 
\begin{equation}
\int_{|\vec r|=a}^R |\vec r|^{-\alpha-1} \ud \vec r = \frac{2\pi}{1-\alpha} \left(R^{1-\alpha} - a^{1-\alpha}\right), 
\end{equation}
we derive the normalized probability distribution of a vortex position inside a cluster as 
\begin{equation}
\tau(\vec r) = \frac{1-\alpha}{2\pi \left(R^{1-\alpha} - a^{1-\alpha}\right)} 
|\vec r|^{-\alpha-1} = \frac{n_\alpha}{N} r^{-\alpha-1}. 
\end{equation}
Here we have introduced the \emph{fractal} mean density, 
\begin{equation}
n_\alpha = 
\frac{N(1-\alpha)}{2\pi \left(R^{1-\alpha} - a^{1-\alpha}\right)}, 
\end{equation}
where \emph{fractal} alludes to the fact that this cluster is self-similar with fractal dimension $1-\alpha$. This $n_\alpha$ is the mean density that is kept fixed when we later take the thermodynamic limit of $N, R \rightarrow \infty$.

A vortex at distance $\vec r$ from the origin of the disk induces a velocity at the origin given by
\begin{equation}\label{eq:v}
\vec \phi(\vec r) = -\frac{\gamma}{2\pi}\frac{\vec r_{\bot}}{r^2},
\end{equation}
where $\gamma = 2\pi\xi c$ is the quantized circulation. The $\bot$ subscript 
denotes the counter-clockwise rotation of a vector with an angle $\pi/2$. Hence, for a configuration of N vortices at positions $\vec r_i$, the velocity induced at the origin is a superposition of the velocity generated by each vortex from Eq.~(\ref{eq:v}), i.e. $\sum_{i=1}^N \vec \phi(\vec r_i)$. 

The velocity at the origin will fluctuate from one cluster configuration to another and, in fact, the probability distribution of a velocity fluctuation equal to $V$ can be calculated by averaging over all possible configurations of clustered vortices that yield a velocity $V$ at origin, namely 
\begin{equation}
  W(\vec V) = \int \left[\prod_{i=1}^N \ud\vec r_i \tau(\vec r_i)\right]\delta\left( V - \sum_{i=1}^N\vec \phi(\vec r_i) \right). \label{eq:Wraw}
\end{equation}
We have assumed that the positions of vortices inside the cluster are uncorrelated, such that the $N$-point configurational distribution factorizes into the N-product of the probability $\tau$ of finding a vortex.  
\section{Formal solution}
In order to decouple the N-dimensional integral from Eq.~(\ref{eq:Wraw}), we Fourier transform the Dirac delta function as
$\delta(\vec x) = \frac{1}{(2\pi)^2}\int e^{-i\vec \rho\cdot \vec x}\ud \vec \rho$, and insert it into Eq.~(\ref{eq:Wraw}), therefore 
\begin{equation}
  W(\vec V) = \frac{1}{(2\pi)^2}\int\ud\vec\rho 
  e^{-i\vec\rho\cdot \vec V}\prod_{i=1}^N\int\ud \vec r_i\tau(\vec r_i)
  e^{i\vec\rho\cdot\vec\phi(\vec r_i)}.
\end{equation}
Upon noticing that the N inner integrals are identical, the $W(\vec V)$ can be simplified to
\begin{align}
  W(\vec V) &= \frac{1}{(2\pi)^2}\int\ud\vec\rho\left( \int\ud\vec r \tau(\vec r)e^{i\vec\rho\cdot\vec\phi(\vec r)} \right)^Ne^{-i\vec \rho\cdot \vec V}, \notag\\
  &= \frac{1}{(2\pi)^2}\int\ud\vec\rho A(\vec \rho)e^{-i\vec \rho\cdot \vec V},
\end{align}
where $A(\vec \rho)$ is the the Fourier transform of $W(\vec V)$ in the velocity's conjugate space $\vec q$ and given by
\begin{align}
  A(\vec \rho) &= \left( \frac{n_\alpha}{N}
  \int_{|\vec r|=a}^R r^{-\alpha-1}e^{i\vec \rho \cdot \vec \phi(\vec r)}\ud \vec r \right)^N \notag \\
  &= \left( 1 - \frac{n_\alpha}{N}\int_{|\vec r| = a}^R r^{-\alpha-1}\left( 1 - e^{i\vec\rho\cdot\vec\phi(\vec r)} \right) \ud \vec r \right)^N. 
\end{align}
Here we made use of the identity $\frac{n_\alpha}{N}\int r^{-\alpha-1}\ud \vec r = 1$, in order to write the integral in a form that converges to the exponential function in the large N-limit, i.e.
\begin{equation}
  \lim_{N\rightarrow \infty}\left( 1 - \frac{x}{N} \right)^N = e^{-x}. 
\end{equation}
This identity is valid as long as $x$ increases less rapidly than $N$. 
Thus, in the thermodynamic limit of large $R$ and $N$, we can
write $A(\vec \rho) = e^{-n_\alpha C(\vec \rho)}$ where 
\begin{equation}
  C(\vec \rho) = \int_{|\vec r|=a}^R r^{-\alpha-1}(1-e^{i\vec \rho\cdot \vec \phi(\vec r)})\ud \vec r, 
\end{equation}
as long as $C(\vec \rho)$ increases less rapidly than $N$.

We now change variables from $\vec r$ to $\vec \phi$. This gives a Jacobi determinant 
\begin{equation}
  \left\| \frac{\partial(\vec r)}{\partial(\vec \phi)}\right\| = -\left(\frac{\gamma}{2\pi}\right)^2\phi^{-4}. 
\end{equation}
Since $|\vec \phi| = \gamma/2\pi r$, we have $r = \gamma/2\pi\phi$. The result is 
\begin{equation}
  C(\vec \rho) = \left( \frac{\gamma}{2\pi} \right)^{1-\alpha}\int_{|\vec \phi|=\gamma/2\pi R}^{\gamma/2\pi a}
  \phi^{\alpha-3}(1-e^{i\vec\rho\cdot \vec \phi})\ud\vec \phi, 
\end{equation}
where the negative sign from the Jacobian is canceled by interchanging the 
limits of integration. Switching to polar coordinates measured relative to the
direction of $\vec \rho$ and rewriting the limits using $\gamma = 2\pi\xi c$, 
\begin{align}
  C(\vec \rho) &= \left( \frac{\gamma}{2\pi} \right)^{1-\alpha}
  \int_{c\xi/R}^{c\xi/a}\phi^{\alpha-2}
  \int_0^{2\pi}\ud\theta \left( 1-e^{i\rho\phi\cos\theta} \right)\ud\phi \notag \\
  &= 2\pi\left( \frac{\gamma}{2\pi} \right)^{1-\alpha}
  \int_{sc}^{s'c}\phi^{\alpha-2}
  \left(1-J_0(\rho\phi)\right)\ud\phi,
\end{align}
where $s = \xi/R$ gives the separation of scales between the coherence length $\xi$
and the system size $R$ and $s' = \xi/a \lesssim 1$ relates the lower cutoff $a$ to the coherence length.
Finally, substituting $x = \rho\phi$, we find 
\begin{align}
  C(\vec \rho) &= 2\pi\left( \frac{\gamma\rho}{2\pi} \right)^{1-\alpha}
  \int_{s\rho c}^{s'\rho c}
  \left[ 1-J_0(x) \right]x^{\alpha-2}\ud x\notag\\
  &= 2\pi\kappa(\rho;s, s') \left( \frac{\gamma\rho}{2\pi} \right)^{1-\alpha},
\end{align}
where $\kappa(\rho;s, s')$ is the dimensionless number given by the integral.
The behavior of this $\kappa$ function in different regimes is the key
to investigating the various regimes of $W(\vec V)$, as we will see in the
next sections.

Summarizing, we have the formal solution 
\begin{align}
  W(\vec V) &= \frac{1}{(2\pi)^2}\int A(\vec \rho)e^{i\vec \rho\cdot \vec V}
  \ud \vec \rho \notag \\
  &= \frac{1}{(2\pi)^2}\int_0^{2\pi}\ud\theta 
  \int_0^\infty \rho e^{-n_\alpha C(\rho)} e^{i\rho V\cos \theta}\ud \rho,
  \label{eq:formsol}
\end{align}
where we introduced polar coordinates with the angle measured relative to
the direction of $\vec V$.
\section{High velocity cutoff}
As we have noted, we have introduced a lower limit $a$ to how close a vortex can
get to the origin. This limits the velocity that a single vortex can induce to 
$V_a = \gamma/2\pi a = s'c$, so we should expect a cutoff in the velocity distribution
around this value.
For velocities larger than $s'c$, values of $\rho$ larger
than $1/s'c$ will tend to cancel out by the oscillating factor $e^{i\rho V}$.
It therefore suffices to consider $\rho < 1/s'c$. In this case, the 
$\kappa(\rho;s,s')$ integral has limits $s'\rho c < 1$ and $s\rho c \ll 1$. 
The lower limit can be taken to zero, while the small upper limit means that
we can expand the Bessel function as $J_0(x) = 1 - x^2/4$, so 
\begin{equation}
  \kappa(\rho;s) = \int_0^{s'\rho c}x^{\alpha}\ud x = 
  \frac{1}{1+\alpha}(\rho c)^{1+\alpha}. 
\end{equation}
Using $s'\rho c = \gamma\rho/2\pi a$, we therefore see that 
\begin{equation}
  C(\vec \rho) = \frac{2\pi}{(1+\alpha)a^{1+\alpha}} \left( \frac{\gamma\rho}{2\pi} \right)^2. 
\end{equation}
Thus the $A(\vec \rho)$ a Gaussian function, which is invariant upon Fourier transformation, and therefore we obtain also a Gaussian distribution for the cutoff-tail of $W(\vec V)$,
\begin{equation}
  W(\vec V)_{V\gg c} = (1+\alpha)\frac{a^{1+\alpha}}{2n_\alpha\gamma^2}
  \exp\left(-(1+\alpha)\frac{\pi a^{1+\alpha}}{2n_\alpha \gamma^2}V^2\right). 
\end{equation}
%
\section{Power-law tail distribution}
We now explore the intermediate scaling regime, where a power-law tail distribution can develop when $sc \ll V < c$. For these velocities, the main contribution to the Fourier transform is when $1/c < \rho \ll 1/sc$.
In the $\kappa(\rho;s)$ integral, this means that the lower limit 
$s\rho c \ll 1$ and can be taken to be zero. The upper limit is larger than $1$, and since the integrand falls off rapidly for $x>1$
we can extend this limit to infinity. Thus, we find that in this regime,
$\kappa(\rho;s) = \kappa$ is a constant and equal to (See appendix \ref{app:kappa})
\begin{align}
  \kappa &= \int_0^{\infty}\left[1-J_0(x)\right]x^{\alpha-2} \ud x \notag\\
  &= - \frac{1}{\pi}\sin \frac{\pi\alpha}{2}\Gamma(\alpha-1)B\left(1-\frac\alpha 2, \frac 1 2 \right).
\end{align}
This means that $C(\rho) \sim \rho^{1-\alpha}$ different from the Gaussian behavior. We explore the consequences of this by studying the Fourier transform integral from Eq.~(\ref{eq:formsol}).

Using the symmetry of the cosine function, we can restrict the polar integration from $0$ to $\pi$ in exchange for a factor of $2$. Changing variables to $t = \cos \theta$ and $z = \rho V$, we find that Eq.~(\ref{eq:formsol}) is equivalent to 
\begin{equation}
 W(\vec V) = \frac{1}{2\pi^2V^2}\int_{-1}^1 \frac{\ud t}{\sqrt{1-t^2}}
  \int_0^\infty z e^{izt}e^{-n_\alpha C(z/V)}\ud z. 
\end{equation}
In order to analyze the high-velocity behavior of this distribution, we would like to expand $e^{-n_\alpha C(z/V)}$ 
in powers of $z/V$ and integrate term by term. However, this interchange of 
limits requires the inner integral to be an analytical function of $t$.
But as any neighborhood of real numbers contains numbers with an imaginary part
of either sign, the $e^{izt}$ factor will cause the integral to blow up
on any neighborhood of $t$.

\begin{figure}[tp]
  \centering
  \includegraphics[width=0.4\textwidth]{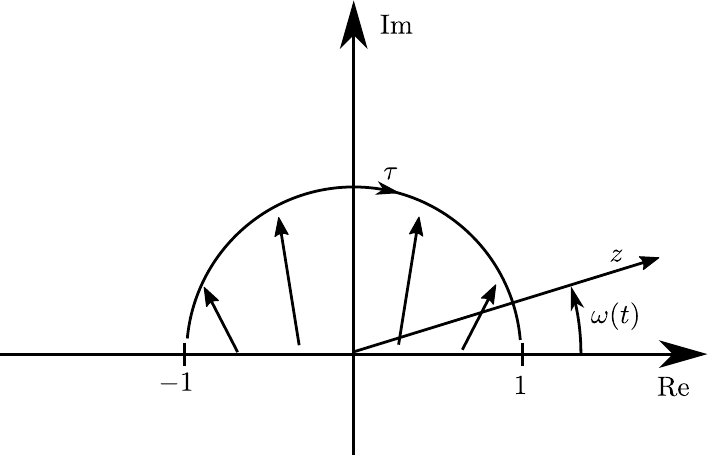}
  \caption{Deforming the contours of integration}
  \label{fig:tcont}
\end{figure}

We can however deform the integration contours in order to ensure that the real part of $izt$ is always negative. The trick is to rotate the ray of the 
$z$ integration by an angle $\omega(t)$ which depends on the argument of $t$. 
In order to avoid a discontinuous change of $\arg t$ from $\pi$ to $0$, we first deform
the $t$-integral to the unit semicircle $S$ in the positive imaginary 
half-plane, as illustrated in figure~\ref{fig:tcont}. Thus $\arg t$ will go continuously
from $\pi$ to $0$.
The exponent $izt$ now has the argument
$\operatorname{arg}(izt) = \pi/2 + \omega(t) + \operatorname{\arg t}$,
and $\omega(t)$ should be chosen so that this is kept between
$\pi/2$ and $3\pi/2$ for the real part to be negative. 
We also need to keep the real part of $C(z/V) \propto z^{1-\alpha}$ negative,
which means that $\operatorname{arg}z^{1-\alpha} = (1-\alpha)\omega(t)$ 
must be kept between $-\pi/2$ and $\pi/2$. These constrains restrain $\omega(t)$ to~
\citep{Chavanis_2000}  
\begin{equation}
  \omega(t) = \frac 1 8 \left( \frac \pi 2 - \arg t \right). 
\end{equation}
We can now expand $e^{n_\alpha C(z/V)}$ in powers of $1/V$ and integrate
term by term: 
\begin{align}
  W(\vec V) = &\frac{1}{2\pi^2V^2}\sum_{n=0}^\infty
  \frac{1}{n!}(-2\pi n_\alpha\kappa)^n
  \left( \frac{\gamma}{2\pi V} \right)^{n(1-\alpha)} \notag\\
  &\times\int_\tau \frac{\ud t}{\sqrt{1-t^2}}\int_{\omega_t} ze^{izt} 
  z^{n(1-\alpha)}\ud z.
\end{align}
After this interchange of limits, we can again rotate the ray of integration 
so that $izt$ is negative real and substitute $y = -izt$, making this integral
a real integral on the positive real axis:
\begin{align}
  W(\vec V) = &-\frac{1}{2\pi^2V^2}\sum_{n=0}^\infty
  \frac{1}{n!}(-2\pi n_\alpha\kappa)^n
  \left( \frac{i\gamma}{2\pi V} \right)^{n(1-\alpha)} \notag\\
  \times&\int_\tau \frac{\ud t}{t^{2+n(1-\alpha)}\sqrt{1-t^2}}
  \int_0^\infty e^{-y} y^{1 + n(1-\alpha)} \ud y \label{eq:series}.
\end{align}
In appendix \ref{app:lambda} we show that the $t$ integral from the $n=0$ term vanishes, 
\begin{equation}
  \int_\tau \frac{\ud t}{t^2\sqrt{1-t^2}} = 0, 
  \label{eq:vanishint}
\end{equation}
while the $n=1$ integral does not, 
\begin{align}
  \lambda &= \int_\tau \frac{\ud t}{t^{3-\alpha}\sqrt{1-t^2}} \notag\\
  &= -\frac i 2 \left( 1+e^{i\pi\alpha} \right)B\left( \frac{3-\alpha}{2}, \frac 1 2 \right),
  \label{eq:lambda}
\end{align}
where $B(a,b)$ is the Beta function.
Thus, recognizing the Gamma function for the $y$-integral, the velocity distribution 
is, to leading order in $1/V$, given by
\begin{align}
  W(\vec V) &= \frac{2\pi n_\alpha\kappa}{2\pi^2V^2}
  \left( \frac{i\gamma}{2\pi V} \right)^{1-\alpha}
  \lambda \Gamma(3-\alpha) \notag\\
  &= \frac{\kappa\lambda}{\pi} \Gamma(3-\alpha) e^{i\pi(1-\alpha)/2} n_\alpha
  \left(\frac{\gamma}{2\pi}\right)^{1-\alpha}V^{\alpha-3}.
\end{align}
In appendix \ref{app:prefactors} we show that the dimensionless prefactors combine to unity, 
\begin{equation}
  \frac{\kappa\lambda}{\pi}\Gamma(3-\alpha)e^{i\pi(1-\alpha)/2} = 1, 
\end{equation}
so that we are left with the simple expression for the tail, 
\begin{equation}
  W(\vec V) = n_\alpha \left( \frac{\gamma}{2\pi} \right)^{1-\alpha}
  V^{\alpha-3}. 
\end{equation}
 
The distribution for the velocity norm picks up a factor $2\pi V$ from the $2D$
measure,
\begin{equation}
  P\left(|\vec V| = V\right) = 2\pi n_\alpha \left( \frac{\gamma}{2\pi} \right)^{1-\alpha}
  V^{\alpha-2}. 
\end{equation}
Taking the limit $\alpha \rightarrow -1$ recovers the familiar $V^{-3}$ 
velocity tail associated with a uniformly random distribution of point 
vortices. On the other hand, by substituting $\alpha = 1/3$ as in the vortex
clustering associated with the inverse energy cascaded, we obtain that the tail distribution develops a $V^{-5/3}$ scaling regime.

\section{Range of the power-law scaling}
\label{sec:pwrange}
The series expansion for $W(\vec V)$ in equation (\ref{eq:series}) contains terms of higher order in $1/V$ which will become increasingly important
for lower velocities $V$. When these terms are included we no longer have a simple power-law scaling, so studying these terms will tell us the
expected scaling range for the high-velocity tail.

The $n=2$ term in equation (\ref{eq:series}) is 
\begin{equation}
  W_2(\vec V) = P_2 n_\alpha^2 \left( \frac{\gamma}{2\pi} \right)^{2-2\alpha}V^{2\alpha-4},
\end{equation}
where the dimensionless prefactor can be shown to equal
\begin{equation}
  P_2 = -4^{1-\alpha}\tan \frac{\pi\alpha}{2}B\left( 1-\frac\alpha 2, \frac 1 2 \right)^2.
\end{equation}
Note that this prefactor is negative when $\alpha > 0$ and positive when $\alpha < 0$. This means that for negative $\alpha$ the distribution will initially increase above the expected power law, before higher-order terms cause the distribution to fall off. Thus the deviation from the power law tail distribution will manifest itself as a ``bulge'' below the scaling regime.
For $\alpha > 0$ the prefactor is negative, so no such bulge appears.

For $\alpha = 0$ the second-order contribution vanishes, so we will need to use the third-order contribution in order to analyze the range of the power-law scaling. The relevant value is 
\begin{equation}
  P_3(\alpha = 0) = -6\pi^2.
\end{equation}

The velocity where the $n=2$ contribution is as important as the $n=1$ term can now be found by solving the equation
\begin{equation}
  |P_2|n_\alpha^2\left( \frac{\gamma}{2\pi} \right)^{2-2\alpha}V^{2\alpha-4} = n_\alpha \left( \frac{\gamma}{2\pi} \right)^{1-\alpha}V^{\alpha-3},
\end{equation}
which gives us a cutoff velocity
\begin{equation}\label{eq:Vcut}
  V_{\text{cut}} = \left( \frac{\gamma}{2\pi} \right)\left( |P_2|n_\alpha \right)^{1/(1-\alpha)}.
\end{equation}
This means that the $V^{\alpha-3}$ can only develop between the lower cutoff velocity $V_{\text{cut}}$ and the speed of sound $c$. Notice that the cutoff value for a given $\alpha$ is entirely controlled by the mean density: the larger the gap between the vortex core size and the mean vortex separation, the wider the range of velocity fluctuations. When $n_\alpha = \xi^{\alpha-1}/|P_2|$ we find that $V_{\text{cut}} = c$, so there is no room for a power law scaling
to develop at densities of this order. Similarly, in order to get a full decade of power-law scaling we need the density to satisfy $n_\alpha < (10\xi)^{\alpha-1}/|P_2|$.

\section{Numerical sampling of clustered vortices}
We check the analytical predictions of the velocity distribution arising from a fractal configuration of $N$-point vortices $\{\vec{r}_i\}_{n=1}^N$, by using the same kind of numerical sampling method as described in~Ref.~\cite{Bradley_2012}.    

The spatial sampling method generates a localized, finite configuration of power-law distributed vortices with respect to their distances from origin, $r_i$, by taking into account the finite vortex core size $a \sim\xi$ and system size $\sim R$. In this case we fix $a = \xi$, so the sampled fractal distribution normalized in the interval bounded by these cutoffs $[\xi,R]$ is
\begin{equation}\label{eq:palpha}
T_\alpha (r) = \frac{1-\alpha}{R^{1-\alpha}-\xi^{1-\alpha}}r^{-\alpha}.
\end{equation}
To sample vortex distances $r_i$ from this distribution, we first generate $N$ random numbers $u_i$ in the unit interval $[0, 1]$. The uniformly distributed random numbers are then mapped onto the set of distances $\{r_i\}$ that follow a fractal distribution given by Eq.~(\ref{eq:palpha}) upon the transformation
\begin{equation}
  r_i = \left[ u_i R^{1-\alpha} + (1-u_i)\xi^{1-\alpha} \right]^{1/(1-\alpha)}.
\end{equation}
The vortex angles $\theta_i$ are assumed to be uniformly distributed between $[0,\pi]$. Therefore the position vector of vortex $i$ is $\vec r_i = r_i(\cos\theta_i\vec e_x + \sin\theta_i\vec e_y)$. 

\begin{figure}[t]
  \centering
  \includegraphics[width=0.49\textwidth]{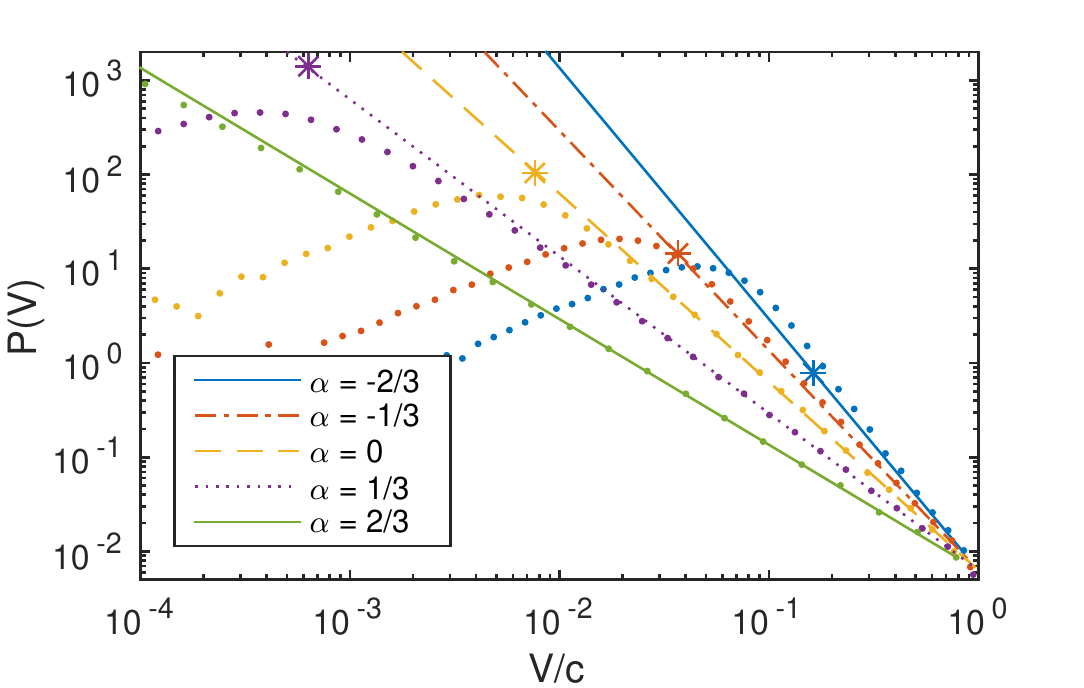}
  \caption{Velocity probability distribution $P(V)$ of a spatially sampled fractal configuration of point vortices with different power-law exponents $\alpha$ ranging from $-2/3$ (steepest power law) to $2/3$ (shallowest power law). The mean density is fixed to $n_\alpha = 10^{-3}\xi^{\alpha-1}$. Straight lines show the corresponding power law $2\pi n_\alpha V^{\alpha-2}$, while the asterisks show the points where the next-order contribution from the power-law expansion is as large as the first-order contribution. Notice that the range of power-law scaling increases with $\alpha$ and the positive second-order contribution at negative $\alpha$, as predicted in section \ref{sec:pwrange}.}
  \label{fig:varypow}
\end{figure}

\begin{figure}[t]
  \centering
  \includegraphics[width=0.49\textwidth]{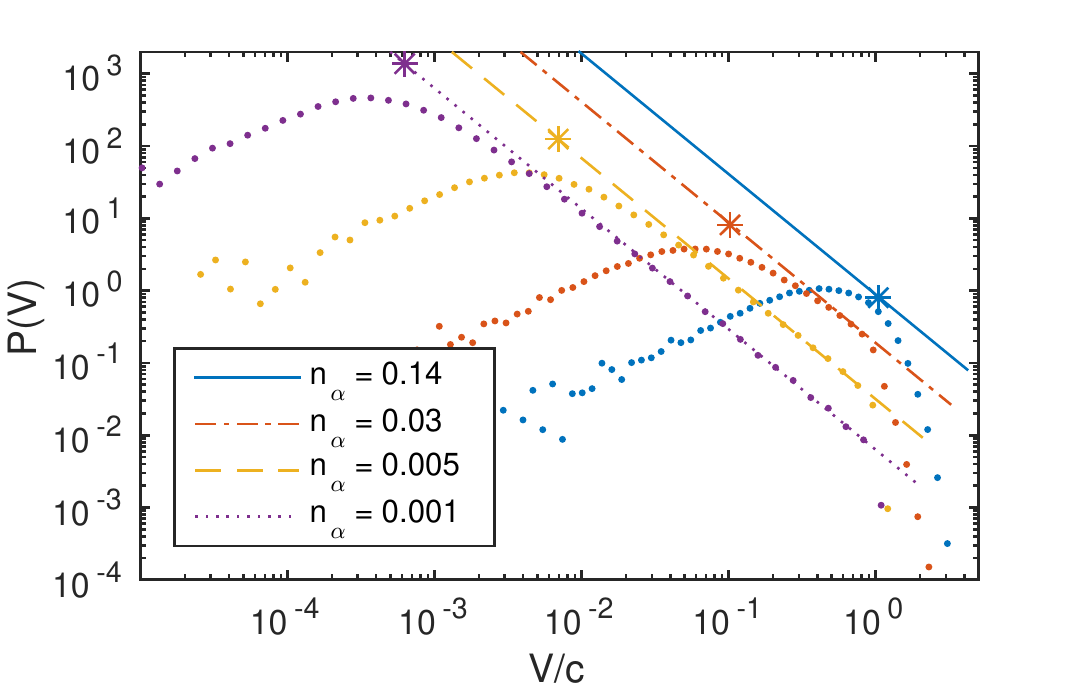}
  \caption{$P(V)$ of a spatially sampled fractal configuration of point vortices with $\alpha = 1/3$ and the mean density $n_\alpha$ varying from $10^{-3}\xi^{\alpha-1}$ to the critical value $0.14\xi^{\alpha-1}$ where the scaling vanishes. Straight lines and asterisks are as in the previous figure. }
  \label{fig:sample}
\end{figure}

\begin{figure}[t]
  \centering
  \includegraphics[width=0.49\textwidth]{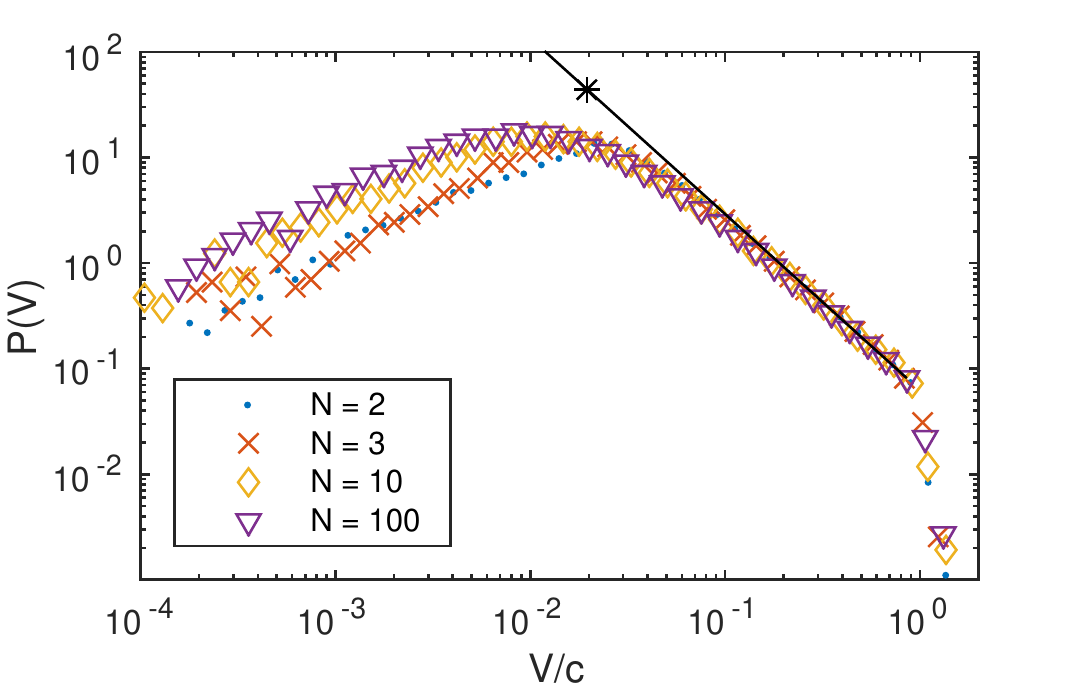}
  \caption{Sampled velocity distributions at different cluster sizes $N$, but at the same density $n_\alpha = 0.01 \xi^{\alpha-1}$ and power-law exponent $\alpha = 1/3$. 
  We notice that the size of the cluster has a minimal effect on the scaling regime, even quite far away from the thermodynamic limit $N,R\rightarrow\infty$ at fixed $n_\alpha$.}
  \label{fig:varyn}
\end{figure}

From the configuration of vortex positions $\{\vec r_i\}_{i=1}^N$, we then compute the velocity induced at the cluster's origin using the superposition principle and Eq.~(\ref{eq:v}), i.e. $\vec v = \sum_{i=1}^N \vec\phi(\vec r_i)$. In Figures~\ref{fig:varypow}, \ref{fig:sample} and \ref{fig:varyn}, we show how the range of power-law scaling of $P(V)$ depends on parameters such as the exponent $\alpha$, the mean vortex density $n_\alpha$, and number of clustered vortices $N$ at a fixed density $n_\alpha$. For a fixed mean density, we notice that the range of the power-law scaling increases with $\alpha$ (see Figure~\ref{fig:varypow}), whereas it remains relatively robust to the number of clustered vortices as shown in Figure~\ref{fig:varyn}. On the other hand, the scaling range extends over more decades in velocity fluctuations as the mean vortex density decreases, as shown in Figure~\ref{fig:sample}, also consistent with the theoretical prediction of lower velocity cutoff $V_{\text{cut}}$ from Eq.~(\ref{eq:Vcut}). This suggests that finite size effects due to small system size and number of clustered vortices have small corrections to the scaling range compared to the dominant effect, which is given by the mean density $n_\alpha$ of clustered vortices.  


\section{Conclusions}
In summary, we have determined the probability distribution of velocity fluctuations arising from fractal configuration of clustered vortices.  

We recover two limit cases that are particularly relevant for turbulent flows. The uniform distribution of vortices that corresponds to $\alpha=-1$ is the stationary configurational probability of a free system of uncorrelated vortices, and used as a proxy to describe 2D turbulent flows with no transfer of energy across scales~\citep{Novikov_1975,Chavanis_2000}. In this case, the induced velocity fluctuations follow the known $V^{-3}$ tail distribution. A similar tail distribution has been observed in 3D superfluid turbulence and attributed to vortex reconnections~\cite{Paoletti_2008}, and also reproduced in 3D simulations of quantum turbulence in Bose-Einstein condensates~\cite{white2010nonclassical,baggaley2011quantum}. Whilst it is plausible that a forward energy cascade in three-dimensions can be described through a uniform tangle of quantized vortices, the inverse energy cascade in 2D turbulence is built on a self-similar distribution of clustered vortices, such that the Kolmogorov spectrum $E(k)\sim k^{-5/3}$ is attributed to $\alpha=1/3$. This spatial self-similarity of vortices induces a different power-law in the tail distribution of the velocity fluctuations, namely as $\sim V^{-5/3}$.

\appendix
\section{Derivation of the $\lambda$ integral}
\label{app:lambda}
We are considering integrals of the type 
\begin{equation}
  I_m = \int_\tau \frac{\ud t}{t^m \sqrt{1-t^2}},
  \label{eq:i_m}
\end{equation}
where $m$ is a possibly fractional power. Of particular interest is the cases $m = 2$ and $m = 3-\alpha$, which
appear as equations (\ref{eq:vanishint}) and (\ref{eq:lambda}), respectively. 

\begin{figure}[t]
  \centering
  \includegraphics[width=0.45\textwidth]{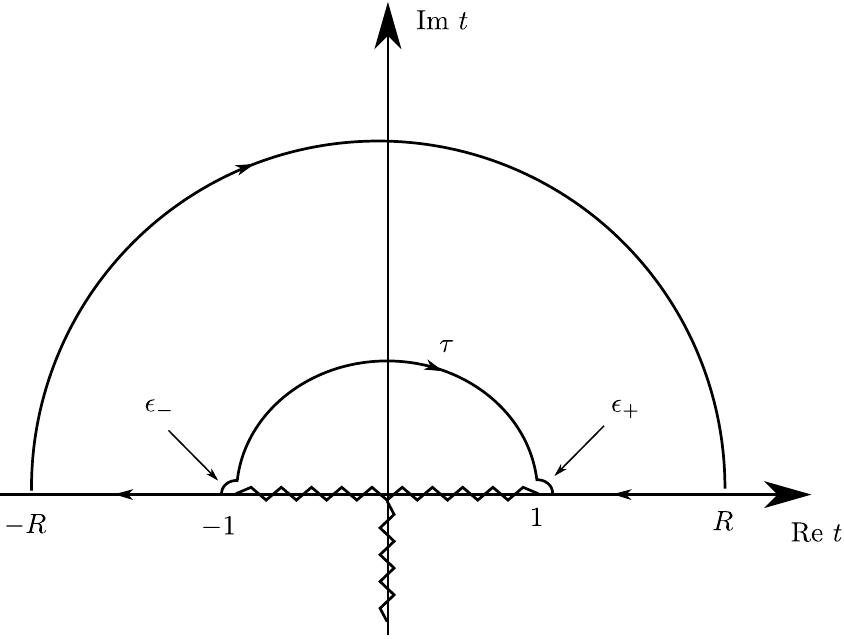}
  \caption{Branch cuts and contours of integration}
  \label{fig:cuts}
\end{figure}

Our strategy will be to deform the integration contour back to the real axis. However, the pole of order $m$ at the origin is likely to
cause problems, which is exactly the reason why we lifted the contour to the complex plane in the first place. We therefore avoid the origin by
enlarging the contour to the intervals $[-1,-\infty]$ and $[1,\infty]$ (see figure \ref{fig:cuts}).

When doing this, we will need to keep any branch cuts out of the way. The fractional power $t^m$ has a branch cut along the negative real axis, 
which we can simply rotate away to the negative imaginary axis by the standard transformation $t^m \rightarrow e^{-i\chi}\left( e^{i\chi/m}t \right)^m$,
with an appropriate choice for $\chi$. This transformation leaves the integral invariant and can thus be left implicit.

For the square root we apply the transformation $\sqrt{1-t^2} = \pm i \sqrt{t^2-1}$ in order to keep the branch cuts
out of the way on the $[-1,1]$ interval. However, we do need to be careful in choosing the sign of the imaginary unit. As $\arg t$ decends from $\pi$ to $0$,
the argument of $1-t^2$ stays in the interval $[-\pi/2, \pi/2]$, not crossing the branch cut of the square root. On the other hand, the argument of
$t^2-1$ crosses the negative real axis at $\arg t = \pi/2$. 

In order to work out the correct sign of the imaginary unit, we calculate the argument of $t^2-1$ using 
the polar form $t = re^{i\theta}$:
\begin{align}
  \arg (t^2-1) &= \arg\left( r^2 e^{2i\theta} - 1 \right) 
  = \arg \left[ r^2 e^{i\theta}\left( e^{i\theta}-e^{-i\theta} \right) \right] \notag\\
  &= \arg \left( 2ir^2\sin\theta e^{i\theta} \right) = \arg e^{i(\theta+\pi/2)} \notag\\
  &= \mathcal{P}(\theta+\pi/2),
\end{align}
where $\mathcal{P}(\theta)$ normalizes the angle to lie in the principal branch interval $(-\pi,\pi]$.
Similarly, the argument of $1-t^2$ is $\mathcal{P}(\theta-\pi/2)$. The principal branch of the square root
halves these angles, so 
\begin{align}
  \arg \frac{\sqrt{1-t^2}}{\sqrt{t^2-1}} = \frac 1 2 \mathcal{P}(\theta-\pi/2) 
  - \frac 1 2 \mathcal{P}\left( \theta+\pi/2 \right).
\end{align}
By checking cases in this expression, we can now verify that 
\begin{equation}
  \frac{\sqrt{1-t^2}}{\sqrt{t^2-1}} = \begin{cases}
	i & \text{when } \pi \le \theta < \pi/2, \\
	-i & \text{when } \pi/2 \le \theta \le 0.
  \end{cases}
\end{equation}
Thus the sign of the imaginary unit changes when the contour crosses the imaginary axis.

The integral can now be written 
\begin{align}
  I_m &= \left( \int_{\epsilon^-} + \int_{-1-\epsilon}^{-R} + \int_{|t|=R} + \int_{R}^{1+\epsilon} + \int_{\epsilon^+}\right)\frac{\ud t}{\pm it^m \sqrt{t^2-1}} \notag \\
  &= I_{\epsilon^-} + I_- + I_R + I_+ + I_{\epsilon^+}.
\end{align}
For the radius $R$ semicircle, we substitue $t = Re^{i\theta}$ and find
\begin{align}
  |I_R| &\le R \int_0^\pi\frac{\ud \theta}{R^m |\sqrt{1-R^2 e^{2i\theta}}|} \notag\\
  &\rightarrow \frac{R}{R^{m+1}}\pi \rightarrow 0,
\end{align}
as long as $m > 0$, which is true for both our cases. Similarly, we find that the $\epsilon$ quarter-circle integrals go like $\sqrt{\epsilon}$, and thus
vanish when $\epsilon\rightarrow 0$. Taking $R \rightarrow \infty$ and $\epsilon \rightarrow 0$, and flipping the limits on the positive integral, we are left with 
\begin{equation}
  I_m = I_- + I_+  = \left( \int_{-1}^{-\infty} - \int_1^\infty \right)\frac{\ud t}{\pm i t^m \sqrt{t^2-1}}. 
\end{equation}

For the negative part we substitute $t \rightarrow -t$, which transforms $\ud t/t^m$ to $-(-1)^{-m}\ud t/t^m$. Inserting the proper
signs for the imaginary unit in each part of the integral, this gives us 
\begin{equation}
  I_m = -i\left( 1 - e^{-i\pi m} \right)\int_1^\infty \frac{\ud t}{t^m \sqrt{t^2-1}}. 
\end{equation}
We now recall the definition of the beta function, 
\begin{equation}
  B(a, b) = \int_0^1 x^{a-1}(1-x)^{b-1}\ud x. 
\end{equation}
Our integral can be transformed into this form with a substitution $x = 1/t^2$, which leads to $\ud t = -\frac{1}{2}x^{-3/2}\ud x$. Thus, 
\begin{align}
  I_m &= -\frac i 2 \left( 1-e^{-i\pi m} \right) \int_0^1 \frac{\ud x}{x^{-m/2}x^{3/2}\sqrt{1/x-1}} \notag\\
  &= -\frac i 2 \left( 1-e^{-i\pi m} \right)\int_0^1 x^{m/2-1}(1-x)^{-1/2} \notag\\
  &= -\frac i 2 \left( 1-e^{-i\pi m} \right)B\left( \frac{m}{2}, \frac 1 2 \right). 
\end{align}
With $m = 2$, or indeed $m$ any even number, this integral vanishes due to $1-e^{-2i\pi} = 0$, which proves equation (\ref{eq:vanishint}).
For the other case, we substitute $m = 3-\alpha$ and find 
\begin{equation}
  \lambda = I_{3-\alpha} = -\frac i 2 \left( 1+e^{i\pi\alpha} \right)B\left( \frac{3-\alpha}{2}, \frac 1 2 \right).
\end{equation}

\section{Derivation of the $\kappa$ integral}
\label{app:kappa}
We are considering the integral 
\begin{equation} \kappa = \int_\epsilon^\infty \left[ 1-J_0(x) \right]x^{\alpha-2}\ud x = I_D - I_J,
\end{equation}
with the understanding that $\epsilon$ should be taken to zero. We would like our results to be valid for $\alpha \in (-1,1)$, which includes
the interesting case $\alpha = 1/3$ and allows us to take the limit $\alpha \rightarrow -1$.
The first term yields a divergence in $\epsilon$, 
\begin{equation}
  I_D = \int_\epsilon^\infty x^{\alpha-2}\ud x = \frac{1}{1-\alpha}\epsilon^{\alpha-1},
  \label{eq:divergence}
\end{equation}
but by series expaning the Bessel function we know that the full integral contains no such divergence. Thus the second term must contain a divergence
which cancels the first one.
Studying this term, we use the integral representation of the Bessel function to write 
\begin{align}
  I_J &= \frac{1}{\pi}\int_\epsilon^\infty \int_0^\pi \cos(x\cos\theta) x^{\alpha-2}\,\ud \theta \ud x \notag \\
  &= \frac{1}{\pi}\int_0^1 \int_{\epsilon}^\infty 2\cos(xt)x^{\alpha-2}\,\ud x \frac{\ud t}{\sqrt{1-t^2}}, \label{eq:IJ}
\end{align}
where we substituted $t = \cos \theta$ and made use of the symmetry of the cosine to halve the integration limits in exchange for a factor of $2$. We now study the 
the integral over $x$, which can be written as
\begin{equation}
  \int_\epsilon^\infty x^{\alpha-2}e^{ixt}\ud x + \int_\epsilon^\infty x^{\alpha-2}e^{-ixt}\ud x = I_x + I_x^*.
\end{equation}
These integrals are suggestive of Gamma function integrals. However, substituting $y = -ixt$ takes the integration contour to the negative imaginary
axis. To avoid this, we first rotate the contour of integration to the positive imaginary axis: 
\begin{equation}
  I_x = \left(\int_{C(\epsilon)} + \int_{i\epsilon}^{iR} - \int_{C(R)}\right)x^{\alpha-2}e^{ixt}\ud x,
\end{equation}
where $C(r)$ is the quarter-circle of radius $r$ going from the real to the positive imaginary axis, and $R$ is to be taken to infinity. The outer quarter-circle integral
vanishes, 
\begin{equation}
  \lim_{R\rightarrow \infty} \int_{C(R)}x^{\alpha-2}e^{ixt}\ud x = 0,
\end{equation}
but on the inner contour we pick up a divergence in $\epsilon$. Substituting $x = \epsilon e^{i\theta}$,
\begin{align}
  I_\epsilon &= \int_{C(\epsilon)}x^{\alpha-2}e^{ixt}\ud x \notag\\
  &= i\epsilon^{\alpha-1}\int_0^{\pi/2}e^{i(\alpha-1)\theta}e^{it\epsilon e^{i\theta}}\ud \theta.
\end{align}
For small $\epsilon$ we can expand the second exponential. In order to keep track of all divergences when $\alpha < 0$ we need to expand to first order,
\begin{align}
  I_\epsilon &= i\epsilon^{\alpha-1}\int_0^{\pi/2}e^{i(\alpha-1)\theta}\left( 1+it\epsilon e^{i\theta} \right)\ud \theta \notag \\
  &= \frac{\epsilon^{\alpha-1}}{\alpha-1}\left( -ie^{i\alpha\pi/2}-1 \right) + i\frac{t\epsilon^\alpha}{\alpha}\left( e^{i\alpha\pi/2}-1 \right) \label{eq:gammadiv}.
\end{align}
Here we assumed $\alpha \neq 0$; the $\alpha = 0$ case is handled below.
In the integral along the imaginary axis we can substitute $y = -ixt$, bringing it to Gamma function form, 
\begin{align}
  \int_{i\epsilon}^{i\infty} x^{\alpha-2}e^{ixt}\ud x &= \left( \frac i t \right)^{\alpha-1}\int_{\epsilon t}^\infty y^{\alpha-2}e^{-y}\ud y \notag \\
  &= -\frac{ie^{i\pi\alpha/2}}{t^{\alpha-1}}\Gamma(\alpha-1, \epsilon t), \label{eq:gammainc}
\end{align}
where $\Gamma(\alpha-1, \epsilon t)$ is the upper incomplete Gamma function. Because $\alpha-1 < 0$, this 
does not actually converge to $\Gamma(\alpha-1)$ when $\epsilon\rightarrow 0$. However, the incomplete gamma function satisfies 
\begin{equation}
  \Gamma(s, x) = \frac{\Gamma(s+1, x)}{s} - \frac{x^s}{s}e^{-x}, 
\end{equation}
as can be verified from an integration by parts. We can make use of this in order to separate out all the diverging terms,
\begin{align}
  \Gamma&(\alpha-1,\epsilon t) = \frac{\Gamma(\alpha, \epsilon t)}{\alpha-1} - \frac{(\epsilon t)^{\alpha-1}}{\alpha-1}e^{-\epsilon t} \notag\\
  &= \frac{\Gamma(\alpha+1, \epsilon t)}{\alpha(\alpha-1)} - \frac{(\epsilon t)^\alpha}{\alpha(\alpha-1)}e^{-\epsilon t} - \frac{(\epsilon t)^{\alpha-1}}{\alpha-1}e^{-\epsilon t}
\end{align}
The first term converges to $\Gamma(\alpha-1)$. To order $\epsilon^\alpha$, we obtain 
\begin{align}
  \Gamma(\alpha-1,\epsilon t) = \Gamma(\alpha-1) - \frac{(\epsilon t)^{\alpha-1}}{\alpha-1} + \frac{(\epsilon t)^\alpha}{\alpha}.
\end{align}
Thus we can see that $I_x$ combines to
\begin{equation}
  I_x = -\frac{\epsilon^{\alpha-1}}{\alpha-1} - it \frac{\epsilon^\alpha}{\alpha} - i\frac{e^{i\pi\alpha/2}}{t^{\alpha-1}}\Gamma(\alpha-1),
\end{equation}
which combines with the complex conjugate to
\begin{equation}
  I_x + I_x^* = 2 \frac{\epsilon^{\alpha-1}}{1-\alpha} + \frac{2}{t^{\alpha-1}}\sin\frac{\pi\alpha}{2}\Gamma(\alpha-1).
  \label{eq:ReIx}
\end{equation}
Inserting this result back into the Bessel function integral in Eq.~(\ref{eq:IJ}), we find
\begin{align}
  I_J &= \frac 2 \pi \int_0^1 \left( \frac{\epsilon^{\alpha-1}}{1-\alpha} + \sin \frac{\pi\alpha}{2}\Gamma(\alpha-1)t^{1-\alpha} \right)\frac{\ud t}{\sqrt{1-t^2}} \notag\\
  &= \frac{\epsilon^{\alpha-1}}{1-\alpha} + \frac 2 \pi \sin \frac{\pi\alpha}{2}\Gamma(\alpha-1)\int_0^1 \frac{t^{1-\alpha}\,\ud t}{\sqrt{1-t^2}}.
\end{align}
Notice the first diverging term here, which will cancel the divergence in Eq.~(\ref{eq:divergence}) exactly.
We can reduce the final integral to a Beta function using a substitution $u = t^2$, 
\begin{align}
  \int_0^1 \frac{t^{1-\alpha}\,\ud t}{\sqrt{1-t^2}} &= \frac 1 2 \int_0^1 u^{-\alpha/2}(1-u)^{-1/2}\ud u \notag\\
  &= \frac 1 2 B\left( 1-\frac \alpha 2, \frac 1 2 \right),
\end{align}
so the final result of the $\kappa$ integral is
\begin{equation}
  \kappa = I_D - I_J = - \frac{1}{\pi}\sin \frac{\pi\alpha}{2}\Gamma(\alpha-1)B\left(1-\frac\alpha 2, \frac 1 2 \right).
  \label{eq:kappagen}
\end{equation}

When $\alpha = 0$, the first-order term from the exponential in Eq.~(\ref{eq:gammadiv}) cancels the $\theta$ dependence, so
\begin{equation}
  I_\epsilon = i\epsilon^{-1}\left( -i-1 \right) - t \frac{\pi}{2}.
\end{equation}
The integral along the imaginary axis in Eq.~(\ref{eq:gammainc}) is simplified to
\begin{equation}
  \int_{i\epsilon}^{i\infty}x^{\alpha-2}e^{ixt}\ud x = -it \Gamma(-1,\epsilon t),
\end{equation}
which diverges logarithmically when $\epsilon \rightarrow 0$. However, as the incomplete gamma function is always real, this
term is purely imaginary and vanishes when we add the complex conjugate. The only contribution is from the $I_\epsilon$ integration,
\begin{equation}
  I_x + I_x^* = I_\epsilon + I_\epsilon^* = 2\epsilon^{-1} - \pi t,
\end{equation}
which we can insert back into Eq.~(\ref{eq:IJ}) to obtain
\begin{equation}
  I_J = \frac{1}{\pi}\int_0^1 \left( 2\epsilon^{-1} - \pi t \right)\frac{\ud t}{\sqrt{1-t^2}}
  = \epsilon^{-1} - 1.
\end{equation}
Canceling the divergence from Eq.~(\ref{eq:divergence}), the result is simply
\begin{equation}
  \kappa_{\alpha=0} = I_D - I_J = 1.
\end{equation}
This is also what one would obtain by taking the limit $\alpha\rightarrow 0$ in the general result from Eq.~(\ref{eq:kappagen}).

\section{Combining the prefactors}
\label{app:prefactors}
The dimensionless prefactor to the power-law tail distribution is
\begin{align}
  P &= \frac{\kappa\lambda}{\pi}\Gamma(3-\alpha)e^{i\pi(1-\alpha)/2} \notag\\
  &= - \frac{1}{\pi^2}\sin \frac{\pi\alpha}{2} C G B,
\end{align}
where $C$ collects the complex number factors, $G$ collects the Gamma functions and $B$ collects the beta functions.
The complex numbers combine to
\begin{equation}
  C = -\frac i 2 \left( 1+e^{i\pi\alpha} \right)e^{i\pi(1-\alpha)/2} 
  = \cos \frac{\pi\alpha}{2},
\end{equation}
so that 
\begin{equation}
  C\sin \frac{\pi\alpha}{2} = \frac 1 2 \sin\pi\alpha.
\end{equation}
For the Gamma functions, we use the well-known properties 
\begin{align}
  \Gamma(x+1) = x\Gamma(x),\quad \Gamma(x)\Gamma(1-x) = \frac{\pi}{\sin\pi x}
\end{align}
in order to write 
\begin{align}
  G &= \Gamma(3-\alpha)\Gamma(\alpha-1) \notag\\
  &= \frac{(2-\alpha)(1-\alpha)}{\alpha-1}\Gamma(1-\alpha)\Gamma(\alpha) \notag\\
  &= -(2-\alpha)\frac{\pi}{\sin\pi\alpha}.
\end{align}
The beta function satisfies
\begin{equation}
  B(a,b) = \frac{\Gamma(a)\Gamma(b)}{\Gamma(a+b)},
\end{equation}
so we can simplify the $B$ factor to
\begin{align}
  B &= B\left( 1 - \frac\alpha 2,\frac 1 2 \right)B\left( \frac{3-\alpha}{2}, \frac 1 2 \right) \notag\\
  &= \frac{\Gamma\left( 1-\frac\alpha 2 \right)\Gamma\left(\frac 1 2\right)}{\Gamma(\frac{3-\alpha}{2})} \frac{\Gamma\left( \frac{3-\alpha}{2} \right)\Gamma\left( \frac 1 2 \right)}{\Gamma\left( 2-\frac\alpha 2 \right)} \notag\\
  &= \pi \frac{\Gamma\left( 1-\frac\alpha 2 \right)}{\left(1-\frac\alpha 2\right)\Gamma\left( 1-\frac\alpha 2 \right)}
  = \frac{2\pi}{2-\alpha},
\end{align}
where we also used that $\Gamma(1/2) = \sqrt{\pi}$.
Combining everything we see that the various factors cancel each other, so in total we have 
\begin{equation}
  P = 1.
\end{equation}

\bibliographystyle{apsrev4-1}
\bibliography{ref2}

\end{document}